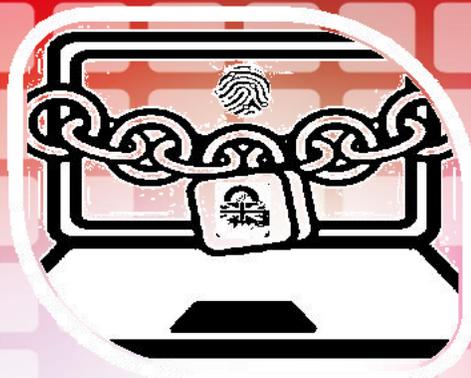

# CYBER THREATS IN FINANCIAL TRANSACTIONS

## Addressing the Dual Challenge of AI and Quantum Computing

**Technical Report**

by the **Cyber Forensics & Threat Investigations Research Community**

**Authored by**

Ahmed M. Elmisery, Mirela Sertovic, Andrew Zayin, Paul Watson

Association of Cyber Forensics and Threat Investigators

**Edited by**

Ahmed M. Elmisery (Director of Training & Outreach)

## Executive Summary

The financial sector faces escalating cyber threats amplified by artificial intelligence (AI) and the looming advent of quantum computing. AI is being weaponized for sophisticated attacks like deepfakes and AI-driven malware, while quantum computing threatens to render current encryption methods obsolete. This report examines these dual threats, analyzes existing legal and regulatory frameworks, and evaluates countermeasures such as quantum cryptography.

AI enhances social engineering and phishing attacks through personalized content and automation, lowers the barrier to entry for cybercriminals, and introduces risks like data poisoning and adversarial AI. Quantum computing, particularly Shor's algorithm, poses a fundamental threat to current encryption standards (RSA and ECC), with estimates suggesting cryptographically relevant quantum computers could emerge within the next 5-30 years. The "harvest now, decrypt later" scenario highlights the urgency of transitioning to quantum-resistant cryptography.

Existing legal frameworks are evolving to address AI in cybercrime, but quantum threats require new initiatives. International cooperation and harmonized regulations are crucial. Quantum Key Distribution (QKD) offers theoretical security but faces practical limitations. Post-Quantum Cryptography (PQC) is a promising alternative, with ongoing standardization efforts.

Recommendations for international regulators include fostering collaboration and information sharing, establishing global standards, supporting research and development in quantum security, harmonizing legal frameworks, promoting cryptographic agility, and raising awareness and education. The financial industry must adopt a proactive and adaptive approach to cybersecurity, investing in research, developing migration plans for quantum-resistant cryptography, and embracing a multi-faceted, collaborative strategy to build a resilient, quantum-safe, and AI-resilient financial ecosystem.

    

## The Dual Threat Landscape in Financial Cybersecurity

The financial sector is experiencing an unprecedented surge in the frequency and sophistication of cyber threats, posing significant risks to the stability and integrity of the global financial system. Financial institutions, entrusted with vast repositories of sensitive data encompassing personal information, transaction records, and proprietary algorithms, have become prime targets for malicious actors [1]. Successful cyberattacks can lead to devastating financial losses for both institutions and individuals, erode public trust in the financial system, and even trigger systemic risks with far-reaching economic consequences [1]. The continuously evolving digital landscape, characterized by increasing interconnectedness and the proliferation of advanced technologies, further amplifies these inherent vulnerabilities.

Within this dynamic environment, the financial industry is confronted by a confluence of emerging risks, where cyberattacks are increasingly being amplified by the rise of artificial intelligence and the looming threat posed by quantum computing [2]. Traditional cybersecurity measures, which have historically served as the bulwark against digital threats, are now facing unprecedented challenges from novel attack vectors and the rapidly advancing capabilities of malicious actors. The cybercrime landscape itself is far from static, with new and evolving threats such as sophisticated tech support scams, intricate investment fraud schemes, and highly manipulative social engineering tactics constantly emerging and escalating in both their prevalence and the magnitude of their financial impact [3]. This fluid and dynamic threat environment underscores the critical need for a proactive and adaptive paradigm in financial cybersecurity, one that transcends reactive defenses and embraces forward-looking frameworks capable of anticipating and effectively neutralizing emerging threats.

The emergence of artificial intelligence (AI) and quantum computing represents a pivotal juncture in the evolution of cyber threats targeting the financial domain [1]. Artificial intelligence has transitioned from being solely a tool for enhancing security measures to becoming a potent instrument in the arsenal of cybercriminals [1]. AI's inherent ability to automate complex tasks, analyze voluminous datasets with remarkable speed, and generate highly realistic and contextually relevant content makes it invaluable for orchestrating sophisticated and widely scalable cyberattacks [4]. Concurrently, quantum computing, while still in its developmental stages, signifies a revolutionary leap in computational power that carries the potential to render currently relied-upon cryptographic standards obsolete [5]. The prospect of quantum computers effectively breaking the widely employed encryption algorithms presents a fundamental and existential threat to the security of financial transactions and the confidentiality of sensitive data [5]. The potential convergence of AI and quantum technologies could further exacerbate these risks, with AI potentially accelerating the development of quantum attack algorithms or significantly enhancing the effectiveness and precision of quantum-enabled cyberattacks. This dual-edged nature of AI and quantum computing in the context of financial cybersecurity presents both opportunities for bolstering defenses and unprecedented threats that demand immediate and comprehensive attention, underpinned by strategic planning and robust regulatory frameworks.

　　　　This report will undertake a detailed examination of the specific cyber threats that AI introduces to financial transactions, encompassing AI-enhanced social engineering techniques, the evolution of AI-driven malware, the risks associated with AI-enabled fraud and identity theft, and the insidious potential of data poisoning and adversarial attacks against AI systems. Furthermore, the report will analyze the impending threat posed by quantum computing to the current cryptographic systems that underpin financial security, with a particular focus on the implications of Shor's algorithm and a critical assessment of the projected timeframe for potential decryption capabilities. The analysis will extend to a comprehensive review of the existing legal and regulatory frameworks at both national and international levels that are intended to combat these emerging threats. The report will also evaluate the feasibility and the inherent legal barriers associated with the adoption of quantum cryptography as a robust and long-term countermeasure. Ultimately, this report aims to propose a set of concrete and actionable steps that international regulatory bodies should proactively undertake to mitigate the multifaceted risks stemming from AI and quantum-enabled cyber threats within the financial sector, thereby fostering a more secure and resilient global financial ecosystem.

## 1. AI-Driven Cyber Threats to Financial Transactions

Artificial intelligence has become a powerful tool for cybercriminals, enabling the development and execution of increasingly sophisticated attacks against financial institutions and their customers [1]. The ability of AI to analyze vast amounts of data, automate complex processes, and generate realistic content has opened up new avenues for malicious activities.

AI algorithms are significantly enhancing the effectiveness and scalability of social engineering attacks and phishing campaigns [7]. Cybercriminals are leveraging AI for in-depth data collection and analysis on potential targets, allowing them to craft highly personalized and convincing attack scenarios [6]. This includes gathering information from social media, professional networking sites, and other publicly available sources to tailor messages that appear legitimate and trustworthy. Generative AI has revolutionized phishing attacks by enabling the creation of highly realistic and contextually relevant emails, SMS messages, and social media outreach [7]. These AI-generated messages often exhibit perfect grammar, natural language, and can convincingly mimic the communication style of trusted individuals or organizations, making them exceptionally difficult to distinguish from genuine communications. Furthermore, AI-powered chatbots can now automate real-time communication in phishing attacks, effectively impersonating customer support agents or other service representatives to gather sensitive personal and financial information [7].

Deepfakes, which are AI-generated manipulated videos, audio recordings, and images, represent an exceptionally potent threat in the context of financial transactions [4]. These highly realistic forgeries can be used for a wide array of malicious purposes, including disseminating misinformation to manipulate markets, impersonating key individuals such as executives or trusted colleagues to authorize fraudulent transactions, and conducting sophisticated social engineering attacks designed to extract sensitive information or induce financial transfers [4]. The widely publicized case of a worker in Hong Kong who was deceived into transferring over $25 million to criminals after an AI-generated video meeting that convincingly emulated the company's CFO and other colleagues serves as a stark illustration of the devastating financial consequences that deepfakes can inflict [8]. Moreover, AI-generated audio can be used to convincingly impersonate individuals' voices, potentially enabling attackers to bypass biometric

authentication systems that rely on voice recognition [8]. This technology can also be deployed in phone scams to deceive individuals into divulging confidential financial details or authorizing fraudulent transactions under the guise of a trusted entity [7]. The capability of AI to generate hyper-realistic and personalized content at scale has significantly amplified the threat of social engineering and phishing attacks, making them considerably more challenging to detect and substantially increasing the likelihood of successful exploitation [4]. Traditional defenses against these types of attacks often rely on identifying inconsistencies or anomalies in language, grammar, or visual cues. However, AI-generated content can effectively overcome these defenses, underscoring the critical need to develop novel detection mechanisms and implement enhanced user awareness training programs that specifically address these advanced threats.

AI is also being integrated into malware to enhance its capabilities and enable it to evade traditional cybersecurity defenses more effectively [6]. AI-powered malware possesses the ability to adapt its behavior in real-time in response to the security environment, making it significantly more challenging for conventional cybersecurity measures to detect, analyze, and ultimately neutralize [6]. Cybercriminals are increasingly utilizing AI to accelerate the development of new and more sophisticated malware variants and to identify and exploit vulnerabilities in target systems with greater efficiency [8]. This automation of the malware development and deployment process can drastically reduce the time window between the discovery of a system vulnerability and its subsequent exploitation by malicious actors. Furthermore, AI can be employed to automate various critical stages of the cyberattack lifecycle, ranging from initial reconnaissance and target selection to lateral movement within a compromised network and the eventual exfiltration of sensitive data [4]. This heightened level of automation enables attackers to launch more widespread, coordinated, and ultimately impactful cyberattacks. AI-enabled ransomware represents another significant concern, as it can leverage AI to conduct in-depth research on potential targets, identify specific system vulnerabilities that can be exploited, and even adapt its data encryption methods over time, thereby making it more resilient to detection by cybersecurity tools and more difficult to counteract [7]. The integration of AI into malware and attack methodologies signifies a substantial escalation in the overall cyber threat landscape, demanding a corresponding and equally advanced evolution in defensive capabilities to effectively counter these emerging threats.

Adversarial AI and data poisoning represent insidious threats to the integrity of AI systems used in financial transactions [7]. Adversarial AI involves the deliberate manipulation of AI systems through carefully crafted inputs that are designed to cause the system to misclassify data, make incorrect predictions, or exhibit other unintended and potentially harmful behaviors [7]. In the context of financial transactions, this could be exploited to bypass fraud detection systems that rely on AI to identify suspicious activities, or to manipulate algorithmic trading platforms to generate illicit profits or cause market disruptions. Data poisoning attacks, on the other hand, directly target the foundational training data that is used to build AI models [4]. By introducing malicious, inaccurate, or biased data into the training process, attackers can compromise the overall integrity and performance of AI applications, leading to the generation of incorrect predictions, the reinforcement of existing biases, or other detrimental outcomes [4]. This is particularly concerning for AI systems that are deployed in critical financial applications such as fraud detection, credit risk assessment, and algorithmic trading, where the accuracy and reliability of the AI's output are paramount. AI models utilized by financial institutions are also susceptible to various other adversarial techniques, including data leakage during the inference phase, where sensitive information about the model or the data it was trained on can be inadvertently revealed, and model extraction attacks, where attackers attempt to reverse-engineer the model's underlying functionality to replicate it or identify and exploit hidden vulnerabilities [15]. The increasing reliance on AI across various facets of financial systems introduces significant new vulnerabilities through these adversarial attacks and data poisoning techniques, which can ultimately undermine the trustworthiness and overall reliability of AI-driven security measures and decision-making processes. Ensuring the continued integrity and robustness of AI models that are integral to financial operations is therefore of paramount importance, given their growing role in performing critical functions within the sector.

AI is significantly enhancing the capabilities of fraudsters in perpetrating financial fraud and identity theft [16]. Generative AI can be employed to create highly realistic fake identities, including the generation of entirely synthetic identities that are not associated with any real person [15]. Fraudsters are also leveraging AI to fabricate sophisticated financial documents, such as bank statements, tax returns, or loan applications, making them appear genuine and substantially more difficult for human reviewers to detect as forgeries [16]. Furthermore, AI can be utilized to automate and scale various identity verification bypass techniques, including the use of deepfakes to circumvent biometric authentication measures such as facial recognition and voice recognition, thereby gaining unauthorized access to accounts and services [8]. While AI algorithms are increasingly being used by financial institutions to analyze vast amounts of financial data to identify patterns and anomalies that may be indicative of fraudulent activity [1], this same capability is also being exploited by criminals to identify potential targets for their schemes and to refine their fraud tactics to evade detection [9]. The widespread availability and increasing sophistication of AI tools are empowering fraudsters with advanced capabilities for creating convincing fake identities, fabricating realistic-looking documents, and effectively bypassing established security measures, leading to a significant increase in both the sophistication and the overall scale of financial fraud and identity theft operations. The ability of AI to generate highly realistic fake credentials and to automate various stages of fraudulent activities poses a substantial challenge to traditional fraud prevention strategies that rely on manual review and rule-based systems.

Illustrative examples of AI-driven cyberattacks in the financial sector are becoming increasingly prevalent, demonstrating the real-world impact of these emerging threats. The widely reported incident involving an employee in Hong Kong who was successfully deceived into transferring a staggering $25 million based on a sophisticated deepfake video conference call convincingly impersonating the company's CFO serves as a stark example of the potentially devastating financial consequences of AI-driven social engineering [8]. Furthermore, cybercriminals have already demonstrated the capability to utilize deepfakes to effectively bypass biometric authentication systems on banking applications in countries like Vietnam and Thailand, allowing them to gain unauthorized access to unsuspecting customers' accounts [8]. The cyberattack targeting the CEO of WPP, which involved a coordinated and multi-faceted approach leveraging a combination of deepfake voice calls and other communication channels, highlights the growing effectiveness of these AI-enhanced attacks [4]. The documented use of generative AI to create entirely fake bank websites specifically designed for the purpose

of harvesting unsuspecting customers' login credentials further illustrates how AI can be readily employed to automate the creation of malicious infrastructure for large-scale phishing campaigns [11]. Additionally, numerous instances have been reported where employees within financial institutions have been successfully tricked into wiring substantial amounts of funds to fraudulent accounts as a direct result of sophisticated AI-driven social engineering attacks, underscoring the tangible and significant impact of these threats on financial organizations [10]. These real-world examples collectively underscore the critical fact that AI-powered cyberattacks are not merely theoretical concerns confined to academic research but are actively being deployed by cybercriminals in the field to inflict significant financial damage on both individuals and organizations, necessitating urgent attention and the implementation of robust and effective countermeasures.

The increasing accessibility and ease of use of AI tools are significantly lowering the barrier to entry for individuals seeking to engage in cybercrime [8]. AI provides cybercriminals with readily available and often user-friendly tools that require considerably less technical expertise to utilize effectively for malicious purposes [8]. This "democratization" of cybercrime effectively allows individuals who may possess limited traditional hacking skills to nonetheless launch sophisticated and impactful cyberattacks. For instance, AI can automate critical tasks such as crafting highly convincing phishing emails that exhibit perfect grammar, natural language, and impeccable spelling, thereby eliminating a key tell-tale indicator that often alerts recipients to fraudulent communications [8]. Moreover, AI tools can assist even novice cybercriminals in writing malicious code or in identifying potential vulnerabilities within target systems, significantly reducing the need for deep and specialized technical knowledge in these traditionally complex areas [8]. The emergence of "Deepfake-as-a-Service" platforms further contributes to this lowering of the barrier by providing relatively easy and affordable access to sophisticated deepfake creation tools for individuals with minimal technical skills, enabling them to generate realistic forgeries for use in scams and social engineering attacks [3]. The development and increasing availability of AI no-code platforms, which can convert natural language instructions directly into functional code, could potentially lead to a significant rise in the number of less technically proficient individuals engaging in cybercrime, as they can leverage these tools to create and deploy malicious software with relative ease [19]. By providing user-friendly interfaces, automating traditionally complex tasks, and reducing the overall technical skill required to execute cyberattacks, AI is significantly lowering the barrier for individuals to engage in malicious online activities, potentially leading to a substantial surge in the number of active attackers and a corresponding increase in the overall volume and sophistication of cyberattacks targeting the financial sector.

## 2. The Quantum Computing Threat to Current Cryptographic Systems

Quantum computing, a revolutionary paradigm in computation, harnesses the principles of quantum mechanics to process information in ways fundamentally different from classical computers [5]. Leveraging phenomena such as superposition and entanglement, quantum computers possess the theoretical capability to solve certain classes of problems with speeds far exceeding those of even the most advanced classical supercomputers. This immense potential, however, casts a long shadow over the security of current cryptographic systems, which underpin the confidentiality and integrity of digital communications and financial transactions [5]. Many of the cryptographic algorithms in widespread use today rely on the computational difficulty of specific mathematical problems, a difficulty that quantum computers could potentially overcome with remarkable efficiency. The ability of quantum computers to rapidly process information could render prevalent encryption methods, including the widely adopted RSA and ECC algorithms, ineffective, potentially exposing vast troves of sensitive data to malicious actors [5]. The revolutionary computational capabilities inherent in quantum computing thus present a fundamental threat to the mathematical foundations upon which current public-key cryptography is built, posing a significant challenge to the security of digital information within the financial sector.

    A key quantum algorithm that poses a direct threat to current cryptography is Shor's algorithm, developed by Peter Shor in 1994 [5]. This algorithm demonstrates the ability to factorize large integers exponentially faster than the most efficient classical algorithms known [5]. The security of the widely used RSA asymmetric encryption algorithm is predicated on the extreme computational difficulty that classical computers face when attempting to factorize the product of two very large prime numbers [22]. Similarly, Elliptic Curve Cryptography (ECC), another prevalent asymmetric encryption

method that offers comparable security to RSA but with smaller key sizes, is also theoretically vulnerable to being broken by quantum computers through variations of Shor's algorithm [5]. The potential for quantum computers to efficiently execute Shor's algorithm implies that the encryption methods that currently safeguard a vast array of critical functions, including financial transactions, the verification of digital signatures, and the establishment of secure communications channels, could become susceptible to decryption by malicious actors possessing quantum computing capabilities [23]. Shor's algorithm, therefore, represents a specific and well-defined quantum threat directed at the core of modern asymmetric cryptography, particularly RSA and ECC, which are indispensable for ensuring the confidentiality and integrity of financial data and the security of financial systems.

While the development of sufficiently powerful quantum computers capable of executing Shor's algorithm at a scale necessary to break current encryption standards is still an ongoing endeavor, experts in the field predict that "cryptographically relevant quantum computers" (CRQCs) with this capability could emerge within the next 5 to 30 years [29]. Some projections even suggest a significant probability of such a development occurring within the next decade [29]. Various estimates exist regarding the precise timeline, with some experts indicating a high likelihood (e.g., 50% or greater) that fundamental public-key cryptography tools will be compromised by the early to mid-2030s [29]. More conservative estimates place this critical juncture closer to the 2040 timeframe or even beyond [31]. The estimated number of physical qubits required to successfully run Shor's algorithm and break commonly used encryption like RSA-2048 is substantial, often cited in the millions [25]. Achieving the necessary level of qubit stability and the implementation of effective error correction mechanisms remain significant technical challenges in the pursuit of building such powerful quantum machines [25]. Despite these ongoing technical hurdles, the rapid and continuous pace of advancements in the field of quantum computing underscores the critical need for proactive preparation for the eventual arrival of CRQCs and the cryptographic vulnerabilities they will introduce [29]. Although the exact timeline for the emergence of cryptographically relevant quantum computers remains a subject of ongoing debate, a strong consensus exists within the scientific and cybersecurity communities that this development poses a substantial threat within the foreseeable future, demanding immediate and strategic action to mitigate its potential impact on the security of financial systems and data.

The potential for quantum computers to break current encryption standards has given rise to the significant "harvest now, decrypt later" (HNDL) threat scenario [12]. This scenario involves malicious actors strategically intercepting and storing vast quantities of encrypted data today, with the explicit intent of decrypting this information at a later date once sufficiently powerful quantum computers become available [12]. This threat is particularly concerning for highly sensitive data that has a long retention period or a lasting value, such as confidential financial records, valuable intellectual property, and classified government secrets [12]. Even data that is currently considered secure due to the strength of its encryption could be retroactively compromised if it falls into the hands of an adversary who can later decrypt it using a quantum computer. Financial institutions, which are often subject to stringent regulatory requirements mandating the retention of various types of data for extended periods, are particularly vulnerable to the risks posed by HNDL attacks [12]. The "harvest now, decrypt later" threat scenario underscores the critical urgency of transitioning to quantum-resistant cryptography. Data encrypted today using currently vulnerable algorithms could be at significant risk of decryption in the future, even if organizations eventually upgrade their systems to quantum-safe methods. This highlights the immediate need for action to safeguard the long-term confidentiality of sensitive financial data.

The impact of quantum computing on financial cybersecurity extends beyond merely compromising the confidentiality of data. Quantum computers also possess the potential to undermine data integrity by enabling the forgery of digital signatures, which are crucial for verifying the authenticity and trustworthiness of electronic documents and transactions [5]. The ability to forge digital signatures could have severe and far-reaching implications for the integrity of financial transactions, the validity of digital contracts, and the overall trustworthiness of digital communications within the financial sector. Furthermore, many blockchain technologies, including various cryptocurrencies and potentially future central bank digital currencies (CBDCs), rely on underlying cryptographic algorithms that are known to be vulnerable to quantum attacks [5]. The capacity of quantum computers to break these fundamental cryptographic algorithms could fundamentally undermine the security and the perceived integrity of these decentralized and distributed systems, potentially leading to significant financial instability and a widespread erosion of public trust in these emerging financial technologies [2]. Additionally, the compromise of authentication mechanisms that rely on currently vulnerable cryptographic algorithms could enable malicious actors to effectively impersonate legitimate systems, manipulate critical transaction records without detection, and gain unauthorized access to highly sensitive financial data and core operational systems [23]. The multifaceted impact of quantum computing on financial cybersecurity, therefore, encompasses not only the threat to data confidentiality but also significant risks to data integrity, the reliability of digital signatures, and the fundamental cryptographic security underpinning both established and emerging financial technologies. This broad spectrum of potential impacts necessitates a comprehensive and proactive approach to preparing for the advent of the quantum era in financial security.

## 3. Legal and Regulatory Frameworks for AI and Quantum-Driven Financial Cybercrime

A foundational layer of national laws exists to address cybercrime and data protection within the financial sector, providing a starting point for tackling AI and quantum-driven threats [19]. In the United States, key legislation includes the Computer Fraud and Abuse Act (CFAA), which criminalizes unauthorized access to computer systems, a provision that is relevant to both AI and quantum-enabled attacks [48]. The Gramm-Leach-Bliley Act (GLBA) mandates that financial institutions implement comprehensive safeguards to protect the security and confidentiality of customer information, a requirement directly applicable to data security in the context of AI-driven threats and the potential for future quantum decryption [52]. The Sarbanes-Oxley (SOX) Act, while primarily focused on financial record accuracy, also includes cybersecurity components to ensure financial institutions address common risks that could impact financial activity, making it relevant to the broader security implications of AI and quantum computing [52]. The Bank Secrecy Act (BSA) requires financial institutions to institute controls that detect and deter money laundering and terrorist financing activities, areas where AI could be exploited by criminals and where quantum vulnerabilities could compromise detection systems [52].

In the European Union, the General Data Protection Regulation (GDPR) imposes

strict rules on the processing and protection of personal data, which is highly relevant to financial institutions operating within the EU or handling the data of EU residents, especially in the context of AI-driven data breaches and the long-term security of personal data against quantum threats [50]. The Digital Operational Resilience Act (DORA) is specifically designed to enhance the operational resilience of the financial sector within the EU, including explicit requirements for managing ICT risks and cryptographic security, with a direct mention of the need to address quantum risks [43]. Other jurisdictions around the globe have enacted similar legislative measures aimed at combating cybercrime and ensuring the protection of sensitive financial data. For example, Singapore, through its Monetary Authority of Singapore (MAS), has issued specific regulations and advisories addressing technology risk management and cybersecurity within financial institutions, including guidance on preparing for quantum computing risks [12]. While a foundational layer of national laws exists to address cybercrime and data protection within the financial sector, providing a crucial starting point for tackling the emerging threats posed by AI and quantum computing, these existing frameworks may require significant updates and supplementation to specifically address the unique characteristics and challenges presented by these rapidly evolving technologies.

Current legal and regulatory frameworks are beginning to acknowledge and address the role of AI in cybercrime, primarily through the application of existing cybersecurity and data protection laws [10]. Many existing cybercrime laws, such as the CFAA, primarily focus on unauthorized access to computer systems and data breaches, but they may not directly address the nuances of AI-specific attack vectors like data poisoning or sophisticated adversarial attacks designed to manipulate AI systems [48]. Data protection regulations like GDPR and GLBA, while emphasizing the critical importance of data security and confidentiality, which are undoubtedly relevant to mitigating the impact of AI-driven attacks, may lack specific guidance on how to secure the AI systems themselves or how to effectively address the unique risks associated with the misuse of AI in the execution of cyberattacks [50]. However, financial regulators are increasingly recognizing the significant implications of AI for cybersecurity within the financial sector and are beginning to incorporate considerations of AI-related risks into their existing risk management frameworks and regulatory guidance [10]. For instance, the New York Department of Financial Services (NYDFS) has already issued specific guidance highlighting the various cybersecurity risks associated with the increasing adoption of AI within financial institutions [10]. Similarly, law enforcement agencies, such as the US Department of Justice (DOJ), are also adapting their strategic approaches to effectively counter cybercrime that involves the use or misuse of AI, acknowledging the technology's potential to be exploited by malicious actors [64]. This growing attention from both regulatory bodies and law enforcement agencies indicates an evolving understanding and a developing response to the emerging challenges posed by AI in the realm of financial cybercrime. While the current legal and regulatory frameworks provide a necessary foundation, there is a clear and increasing recognition of the need for more specific guidance and potentially new regulations that are specifically tailored to address the unique and rapidly evolving risks posed by AI in this critical context.

Several jurisdictions are in the process of exploring or actively implementing regulations that specifically target the use of artificial intelligence within the financial sector [63]. In the United States, for example, some states are considering new legislation that would address the potential disparate impact of AI-aided decisions in areas such as credit and insurance underwriting [65]. Furthermore, the state of New York has proposed specific requirements for banks to notify loan applicants about the use of AI in their lending decisions and to disclose the types of data that these AI tools utilize [65]. At the federal level in the US, the Department of the Treasury has released a series of reports outlining key recommendations for effectively managing AI-specific cybersecurity risks within financial services, suggesting several areas where future regulatory guidance, standards, and even new laws are likely to emerge in the near future, which will significantly impact the permissible use cases of generative AI for financial institutions and the compliance measures they will be required to implement [63]. The European Union's comprehensive AI Act aims to establish a pioneering framework for regulating the development and deployment of artificial intelligence across various sectors, including the financial industry, with potential implications for its use in areas such as cybersecurity and fraud detection [68]. Beyond these major regulatory initiatives, other countries like Nigeria and India are also actively developing their own national AI strategies and guidelines, which are expected to significantly influence the future regulation of AI in their respective financial sectors, with cybersecurity being recognized as a critical consideration within these broader policy frameworks [68]. The emergence of these specific national regulations and policy initiatives that are directly focused on the use of AI in finance clearly signals a proactive and forward-thinking approach to governing the development and deployment of this transformative technology within this highly regulated sector, with a strong emphasis on addressing the associated cybersecurity risks.

Existing international treaties and agreements on cybercrime provide a crucial foundation for fostering cooperation in the global fight against digital threats, including those potentially driven by artificial intelligence [19]. The Budapest Convention on Cybercrime, as the first international treaty specifically focused on addressing cybercrime, aims to harmonize national laws across signatory nations, improve international investigative methods, and promote effective cooperation between countries in combating cyber threats [19]. The broad scope of this convention, which addresses various forms of cybercrime, could potentially be applicable to AI-driven attacks, providing a legal framework for international cooperation in investigations and prosecutions. Similarly, the United Nations Convention against Transnational Organized Crime includes provisions that can be applied to cybercrime, particularly in cases where such activities are carried out by organized criminal groups operating across national borders, which could increasingly involve the use of sophisticated AI tools in the future [51]. International law enforcement organizations, such as Interpol and Europol, play a vital role in coordinating global efforts to combat cybercrime that transcends national boundaries, including facilitating the sharing of critical intelligence between member countries and conducting joint operations targeting cybercriminals [50]. Looking ahead, the Council of Europe is actively developing a new convention specifically on artificial intelligence, which will address a wide range of issues related to the governance, accountability, and risk assessment of AI technologies. This new convention could have significant implications for the use of AI in cybersecurity, potentially establishing international norms and standards for its safe and responsible deployment [68]. While these existing international treaties and agreements provide a valuable foundation for cooperation in combating cybercrime, including potentially AI-driven attacks, the rapid and continuous evolution of AI technologies may necessitate the development of further international agreements or specific protocols to more effectively address these novel and increasingly sophisticated threats.

Current legal frameworks, while evolving, primarily focus on addressing cyber threats originating from classical computing environments and may not yet adequately address the specific and unique risks posed by the advent of quantum computers, particularly the potential for breaking currently used encryption through algorithms like Shor's algorithm. However, the increasing global awareness of the looming quantum threat to cybersecurity is beginning to spur new legislative and regulatory initiatives aimed at preparing for this future challenge. For instance, the United States has enacted the Quantum Computing Cybersecurity Preparedness Act, which mandates federal government agencies to develop and implement plans for migrating their information technology systems to cryptographic methods that are resistant to attacks from both quantum computers and classical computers [12]. While the initial focus of this legislation is on securing government systems, it serves as an important precedent and can encourage similar proactive measures within the private sector, including the highly regulated financial industry. Furthermore, regulatory bodies in various jurisdictions are beginning to issue guidance and advisories specifically addressing the cybersecurity risks associated with quantum computing. The Monetary Authority of Singapore (MAS), for example, has issued advisories to financial institutions on the importance of understanding and addressing the potential threats posed by quantum technology to their cryptographic security [12]. Similarly, the European Union's Digital Operational Resilience Act (DORA) explicitly mentions quantum risks within its regulatory framework, specifically in the context of ICT risk management for financial entities operating within the EU [43]. These emerging legal and regulatory responses indicate a growing recognition of the need to proactively prepare for and mitigate the unique risks that quantum computing poses to current cryptographic systems and the overall security of the financial sector. While the legal and regulatory landscape for quantum-enabled cybercrime is still in its early stages of development, the increasing momentum behind these initiatives suggests a proactive approach to addressing this significant future threat.

    

## 4. Quantum Cryptography as a Countermeasure

Quantum Key Distribution (QKD) represents a groundbreaking cryptographic technique that leverages the fundamental principles of quantum mechanics to establish an exceptionally secure communication channel specifically for the exchange of encryption keys [73]. Unlike traditional classical cryptography, which relies on the inherent computational difficulty of solving certain mathematical problems, the security of QKD is deeply rooted in the fundamental laws of physics themselves [74]. The core principle underpinning QKD involves the transmission of individual particles of light, known as photons, in specific quantum states referred to as qubits, to encode the individual bits of the encryption key [20]. A crucial aspect of QKD is that any attempt by an unauthorized third party, or eavesdropper, to intercept or measure these transmitted qubits will inevitably disturb their delicate quantum state. This disturbance is not subtle; it leaves a detectable trace that can be identified by the legitimate communicating parties [74]. This inherent property allows the two parties engaged in communication to generate and securely share a secret cryptographic key that is provably secure against any and all eavesdropping attempts, regardless of the computational power that the eavesdropper might possess, including that of a future quantum computer [74]. Once this securely established key has been exchanged, it can then be utilized in conjunction with a robust symmetric encryption algorithm, such as the Advanced Encryption Standard (AES), to encrypt and subsequently decrypt the actual data being transmitted between the parties, thereby achieving a very high and theoretically unbreakable level of security for the communication [74]. QKD, therefore, offers a revolutionary approach to secure communication by directly exploiting the fundamental principles of quantum mechanics to guarantee the security of the critical key exchange process, presenting a potential long-term and robust solution against the significant threat posed by future quantum computers to currently used classical cryptographic methods.

The feasibility of employing Quantum Key Distribution (QKD) for securing financial transactions presents both compelling advantages and significant challenges. One of the primary advantages of QKD is its provision of "unconditional security" [78]. This means that the security offered by QKD is not based on any assumptions about the computational limitations of potential attackers and holds true even against adversaries with theoretically unlimited computational power, including advanced quantum computers. Furthermore, QKD inherently allows for the real-time detection of any and all eavesdropping attempts [77]. Due to the fundamental principles of quantum mechanics, any attempt to intercept the quantum key transmission will inevitably introduce a detectable disturbance in the quantum state of the photons, alerting the legitimate communicating parties to the presence of an eavesdropper. QKD can also offer what is known as "retrospective decryption protection" [78]. This crucial feature ensures that even if encrypted messages are intercepted and stored today by malicious actors, they cannot be decrypted in the future by quantum computers if the original encryption keys were securely established and exchanged using QKD. In essence, QKD provides a robust mechanism for establishing cryptographic keys with a level of security that is theoretically unbreakable, offering a substantial advantage for safeguarding highly sensitive financial data and the integrity of critical financial transactions [74]. The primary strengths of QKD, therefore, lie in its ability to provide a level of security that is independent of advancements in computational power and its inherent capability to detect any unauthorized access or tampering during the key exchange process. These advantages make QKD a particularly attractive solution for securing high-value and long-term sensitive financial information where the consequences of a security breach are exceptionally severe.

Despite its strong theoretical security underpinnings, the practical implementation of QKD for the purpose of securing financial transactions faces a number of significant challenges that need to be addressed for its widespread adoption. The cost associated with implementing QKD systems can be quite substantial [78]. This involves the acquisition of specialized and often expensive hardware components, such as highly sensitive single-photon sources and detectors, as well as the establishment of dedicated quantum channels for transmitting the photons, which may involve the deployment of specialized fiber-optic networks or free-space optical links. The initial capital investment in building this infrastructure, along with the ongoing costs of maintenance and operation, can be a significant financial barrier. Furthermore, the effective range over which QKD can be reliably implemented is currently limited by the inherent loss of photons in the transmission medium, particularly in standard fiber-optic cables [78]. While ongoing technological advancements are continually pushing these boundaries, achieving long-distance QKD, especially over continental or global scales, still presents significant technical hurdles and may necessitate the use of trusted intermediate relays or even satellite-based quantum communication systems [74]. QKD systems themselves are technically complex to deploy, configure, and operate, requiring a high degree of specialized expertise in areas such as quantum mechanics, cybersecurity principles, and advanced networking technologies [78]. Integrating these complex QKD systems seamlessly with the existing and often legacy infrastructure within financial institutions can also pose substantial technical challenges. A notable limitation of standard QKD protocols is that they do not inherently provide a mechanism for authenticating the identities of the communicating parties [78]. This means that while the key exchange process itself might be demonstrably secure, there remains a potential risk of a man-in-the-middle attack if the identities of the sender and the receiver are not rigorously verified through independent means, such as classical authentication protocols [77]. Finally, QKD systems, despite their quantum-level security for key exchange, can still be vulnerable to physical layer attacks that target the actual hardware and the specific implementation of the system, often referred to as side-channel attacks [78]. Therefore, ensuring the overall security of the entire QKD system, and not just the quantum key exchange protocol itself, is absolutely crucial for its effective deployment in securing financial transactions. These limitations, encompassing cost, range, technical complexity, the need for robust authentication, and physical layer vulnerabilities, currently restrict the overall scalability and broad applicability of QKD for securing all types of financial transactions and across all sizes and types of financial institutions.

　　　　The adoption of QKD in the financial sector also faces potential legal and regulatory barriers due to the nascent state of the regulatory landscape for quantum cryptography [80]. There is currently a lack of well-established and universally accepted standards and clear regulatory guidelines specifically governing the deployment and use of QKD technology in regulated industries like finance. This regulatory uncertainty can create significant barriers to widespread adoption, particularly within the highly regulated financial sector where compliance with established rules and standards is paramount. Issues related to the standardization of QKD technology, its various protocols, and its interoperability with existing systems need to be thoroughly addressed to ensure seamless integration and to facilitate broader deployment across the industry

 

[80]. The absence of consistent standards could potentially hinder investment in QKD solutions and create significant compatibility problems between different systems and institutions.

Existing regulatory frameworks, which were primarily designed with classical cryptographic methods in mind, may need to be adapted or supplemented to adequately accommodate the unique characteristics of QKD, such as its fundamental reliance on principles of physics and its inherent ability to detect any eavesdropping attempts [81]. Current regulations might not be well-suited to address these novel aspects of quantum cryptography, potentially leading to compliance challenges and hindering adoption. Furthermore, concerns about the potential impact of widespread QKD deployment on national security interests and the ability of law enforcement agencies to access communications under lawful warrants might also lead to regulatory hurdles in certain jurisdictions [79]. Striking a delicate balance between the imperative for highly secure communication within the financial sector and other legitimate societal interests can be a complex regulatory challenge. The nascent state of legal and regulatory frameworks for quantum cryptography, including the current lack of comprehensive standards and clear guidelines, poses a potential and significant barrier to the widespread adoption of QKD within the financial sector. This underscores the critical need for proactive and ongoing engagement between regulatory bodies and the financial industry to establish a supportive yet secure regulatory environment that fosters responsible innovation and the adoption of quantum-safe technologies.

Post-Quantum Cryptography (PQC) represents a critical alternative or potentially complementary approach to QKD for securing financial transactions in the evolving quantum era [73]. PQC encompasses a diverse class of cryptographic algorithms that are specifically designed to be secure against the computational power of both classical computers that exist today and the anticipated capabilities of future quantum computers [73]. These algorithms are based on a variety of mathematical problems that are currently believed to be computationally hard for both types of computers to solve efficiently. Recognizing the urgent need for quantum-safe cryptographic solutions, the National Institute of Standards and Technology (NIST) in the United States has been spearheading a significant international effort to standardize PQC algorithms. In 2024, NIST announced the first set of selected algorithms for both general encryption and digital signatures that are deemed to be resistant to quantum attacks [30]. These standardized PQC algorithms are crucial for providing clear guidance and a robust framework for the transition to quantum-safe cryptography across various sectors, including the highly regulated financial industry. One of the key advantages of PQC algorithms is that they can be implemented on existing classical computing infrastructure without requiring the deployment of entirely new and specialized quantum hardware like QKD networks [84]. This inherent compatibility with current systems makes the adoption of PQC potentially more straightforward and less disruptive for many financial institutions. Some leading financial institutions are already considering and even implementing a "dual quantum-safe strategy" that strategically combines the strengths of both QKD and PQC [86]. In this approach, QKD might be utilized for ultra-secure key exchange in specific, highly critical communication links where the absolute highest level of security is required, while PQC algorithms are employed for broader encryption needs across the wider financial ecosystem. While PQC is widely regarded as a very promising approach for achieving quantum-safe security, it is important to note that ongoing research and rigorous analysis are absolutely crucial to continuously assess and ensure the long-term security of these algorithms against potential future breakthroughs in both classical and quantum computing capabilities [78]. Post-quantum cryptography represents a viable and potentially more readily deployable alternative or complement to QKD for securing financial transactions in the quantum era. The ongoing standardization efforts provide a clear and actionable path for adoption within the financial sector, although continuous vigilance and dedicated research remain necessary to maintain its long-term effectiveness against evolving threats.

## 5. Recommendations for International Regulators to Mitigate AI and Quantum-Enabled Cyber Threats

International regulatory bodies have a crucial role to play in fostering a secure and resilient financial ecosystem in the face of emerging cyber threats from AI and quantum computing. To effectively address these challenges, a coordinated and multi-faceted approach is required.

It is imperative for international regulatory bodies to establish robust platforms and mechanisms that foster enhanced collaboration and facilitate the seamless sharing of critical information regarding the evolving cyber threats posed by both AI and quantum computing within the financial sector [36]. This should include the regular and timely sharing of threat intelligence, the dissemination of best practices for effective mitigation strategies, and the open exchange of insights into the various regulatory approaches being adopted across different jurisdictions. Encouraging the formation of international working groups and specialized forums that bring together a diverse range of key stakeholders, including regulators from various countries, representatives from financial institutions of all sizes, leading technology providers, and renowned academic experts, will be crucial for collectively addressing these complex and rapidly evolving challenges [36]. Furthermore, efforts should be directed towards promoting the development of secure and reliable channels for cross-border information exchange specifically focused on cyber incidents and the tactics, techniques, and procedures of threat actors who are leveraging AI and quantum technologies to perpetrate financial crime. Supporting international initiatives that facilitate the sharing of anonymized threat data will be essential for enhancing the detection and prevention capabilities across the entire global financial ecosystem. Enhanced international collaboration and robust information sharing are paramount for building a unified and resilient global defense against the borderless threats posed by AI and quantum computing in the financial sector, ultimately enabling a more effective and coordinated response to these emerging challenges.

International regulatory bodies should actively work towards the establishment of globally recognized standards and the promotion of best practices for ensuring both AI and quantum security within the financial sector [36]. This includes developing specific guidelines for the secure development and responsible deployment of AI in financial applications, with a particular emphasis on mitigating AI-driven cyber threats such as the creation and use of sophisticated deepfakes, the insidious risks of data poisoning attacks, and the potential for adversarial manipulation of AI systems. Furthermore, it is crucial to collaborate closely with international standards organizations, such as the International Organization for Standardization (ISO) and the US National Institute of Standards and Technology (NIST), to develop and actively promote the widespread adoption of post-quantum cryptographic standards within the financial industry [36]. Developing comprehensive guidelines for financial institutions on how to conduct thorough risk assessments that specifically address the unique vulnerabilities associated with both AI and quantum technologies is also essential [12]. Additionally, establishing best practices for achieving and maintaining cryptographic agility within financial institutions will enable them to rapidly and efficiently adapt their cryptographic infrastructure in response to the emergence of new threats or the evolution of security standards [45]. The development and promotion of standardized international guidelines and best practices are absolutely essential for providing a clear, consistent, and actionable framework for financial institutions worldwide to effectively address the novel and complex security challenges presented by AI and quantum computing, ultimately fostering a more secure and resilient global financial system.

International regulators should actively advocate for and provide strong support for increased funding and resources dedicated to research and development in the critical areas of quantum-resistant cryptography and other quantum security technologies, including quantum key distribution (QKD), through both government-led initiatives and strategic public-private partnerships [27]. It is also crucial to encourage and facilitate robust collaboration between academic institutions that are at the forefront of quantum research, specialized research organizations, and the financial industry itself to accelerate the pace of development and rigorous testing of effective quantum-safe security solutions. Furthermore, efforts should be made to promote the open sharing of research findings, technological advancements, and emerging best practices in the field of quantum security across the international community. Supporting the establishment of international research consortia that are specifically focused on addressing the multifaceted cybersecurity implications of both AI and quantum computing within the financial sector will also be invaluable in fostering innovation and the development of effective countermeasures [95]. Proactive promotion and robust support of sustained research and development in quantum security technologies are absolutely crucial for staying ahead of the rapidly evolving threat landscape and ensuring the long-term availability of effective and robust countermeasures for the financial sector in the face of these transformative technological advancements.

It is essential for international regulatory bodies to work collaboratively towards the harmonization of legal frameworks and regulations related to cybercrime, particularly concerning the potential misuse of artificial intelligence and the exploitation of vulnerabilities associated with quantum computing within the financial sector [43]. Consideration should be given to the potential need for specific legal frameworks that clearly define and effectively criminalize AI-enabled cyberattacks, such as the creation and malicious use of deepfakes for financial fraud or the intentional manipulation of financial AI systems for illicit purposes. Furthermore, it will be necessary to develop legal frameworks that specifically address the unique challenges associated with investigating and prosecuting cybercrimes that involve quantum technologies, including complex issues related to jurisdictional boundaries and the collection of digital evidence in a quantum context. It is also crucial to ensure that existing data protection and privacy regulations are carefully reviewed and adapted to adequately address the specific implications of both AI and quantum computing, such as the enhanced potential for AI to facilitate large-scale data breaches and the future decryption of currently protected sensitive data by quantum computers. The establishment of clear and harmonized legal and regulatory frameworks at the international level is absolutely essential for effectively deterring, detecting, and ultimately prosecuting cybercriminals who seek to leverage the power of AI and quantum computing to target the financial sector. Consistent and harmonized legal frameworks across jurisdictions will help to close potential loopholes that criminals could exploit and ensure that they can be held fully accountable for their actions, regardless of their geographic location.

International regulatory bodies have a critical role to play in actively supporting initiatives that promote cryptographic agility and facilitate the necessary transition to post-quantum cryptography within the financial sector [31]. This includes issuing clear and comprehensive guidance to financial institutions and establishing realistic and achievable timelines for the industry-wide transition to post-quantum cryptographic methods. This may also involve setting specific target dates for the eventual phasing out of currently used cryptographic algorithms that are known to be vulnerable to quantum attacks. Furthermore, regulators should actively encourage and facilitate the development of robust tools and effective frameworks that enable cryptographic agility within financial institutions, allowing them to rapidly and efficiently update their cryptographic systems as new threats emerge or as security standards evolve. Providing readily accessible resources and dedicated support to assist financial institutions, particularly smaller organizations that may lack extensive in-house expertise, in understanding the complexities of post-quantum cryptography and implementing these new security measures will be crucial. Promoting the adoption of hybrid cryptographic approaches, which strategically combine the security of classical and post-quantum algorithms during the transition period, can also provide an added layer of security and facilitate a smoother migration process [5]. International regulators have a crucial role in driving the essential transition to a quantum-safe cryptographic environment within the financial sector by providing clear direction, establishing realistic timelines, and offering the necessary support for the widespread adoption of post-quantum cryptography and the development of robust cryptographic agility capabilities. Regulatory leadership in this area is necessary to ensure a timely, coordinated, and ultimately successful transition across the entire global financial industry.

It is of paramount importance for international regulatory bodies to develop and actively support initiatives aimed at raising awareness and providing comprehensive education on the emerging cyber threats posed by AI and quantum computing for both financial institutions and the consumers they serve [8]. This includes the development and wide dissemination of accessible educational materials and the implementation of effective public awareness campaigns designed to inform financial institutions, their employees at all levels, and the general public about the specific risks associated with AI and quantum-enabled cyber threats. Furthermore, it is crucial to support the development and implementation of specialized training programs for cybersecurity professionals working within the financial sector to equip them with the advanced knowledge and specialized skills necessary to effectively address these rapidly evolving threats, including in-depth knowledge of AI security principles and the intricacies of quantum-safe cryptography [8]. Financial institutions should be strongly encouraged to incorporate comprehensive training on the identification and mitigation of AI-driven social engineering techniques and the detection of sophisticated deepfakes into their mandatory cybersecurity awareness programs for all employees [8]. Additionally, promoting public awareness initiatives aimed at educating consumers about the increasing sophistication of financial scams that are enabled by AI technologies and the potential future risks that may arise from quantum computing is essential for building a more informed and resilient user base. Raising awareness and providing targeted education are critical for fostering a strong culture of cybersecurity vigilance both within financial institutions and among consumers, ultimately empowering them to better understand and effectively defend against the evolving and increasingly complex threats emanating from AI and quantum computing.

## Conclusion: Towards a Quantum-Safe and AI-Resilient Financial Ecosystem

The financial sector is currently navigating a rapidly transforming threat landscape where artificial intelligence and quantum computing are emerging as dominant forces. AI is already being actively exploited by cybercriminals to orchestrate more sophisticated and widely distributed attacks, while quantum computing looms as a future threat capable of undermining the fundamental cryptographic protections that currently safeguard the financial system. Current cryptographic systems, particularly the widely used asymmetric encryption algorithms such as RSA and ECC, are vulnerable to being broken by sufficiently advanced quantum computers within a timeframe that is no longer considered distant. This necessitates a proactive and comprehensive shift towards the adoption and implementation of quantum-resistant cryptographic methods. Legal and regulatory frameworks at both national and international levels are beginning to recognize and address the significant implications of AI and the potential of quantum computing for enabling cybercrime within the financial sector. However, further development, refinement, and, importantly, international harmonization of these frameworks are essential to effectively deter, detect, and ultimately counter these increasingly sophisticated threats. Quantum Key Distribution (QKD) offers a theoretically robust and secure method for cryptographic key exchange, but its practical limitations in terms of cost, range, and technical complexity, coupled with the current lack of comprehensive regulatory frameworks, currently hinder its widespread adoption across the financial industry. Post-quantum cryptography (PQC) presents itself as a promising and potentially more readily deployable alternative or a valuable complementary approach to QKD, with significant standardization efforts currently underway to guide its implementation.

The financial industry must embrace a proactive and adaptive approach to cybersecurity, moving decisively beyond traditional reactive measures to effectively anticipate and strategically mitigate the emerging threats posed by both AI and quantum computing. This requires a continuous and vigilant monitoring of the evolving threat landscape, sustained and strategic investment in cutting-edge research and development of security technologies, and a firm commitment to building robust cryptographic agility within financial institutions. It is imperative that financial institutions begin preparing now for the potential and eventual impacts of quantum computing by developing and diligently implementing comprehensive strategies for transitioning their critical systems to quantum-resistant cryptography. This crucial process includes conducting thorough and detailed risk assessments that specifically address quantum vulnerabilities, meticulously inventorying all cryptographic assets currently in use, and developing clear and actionable migration plans.

The future of financial cybersecurity in the rapidly approaching age of AI and quantum computing will undoubtedly demand a multi-faceted and highly collaborative approach. This will necessitate not only the continuous development and widespread deployment of increasingly advanced security technologies and countermeasures but also the establishment of robust, adaptable, and internationally harmonized legal and regulatory frameworks. Furthermore, fostering strong international cooperation and promoting a culture of comprehensive education and awareness throughout the financial industry and among consumers will be paramount. By taking proactive and decisive steps today, international regulatory bodies and the financial industry can work in concert to build a resilient quantum-safe and AI-resilient financial ecosystem that can continue to operate with integrity, maintain the trust of the public, and effectively safeguard financial assets in the face of these transformative and potentially disruptive technological advancements.

| Source | Year of Estimate | Estimated Timeframe for Breaking RSA-2048 | Conditions Mentioned |
|---|---|---|---|
| National Institute of Standards and Technology (NIST) | 2024 | As soon as 2030 | First breaches |
| Dr. Michele Mosca (University of Waterloo) | 2024 | 50% chance by 2031 | |
| Poll of Experts | 2024 | Likely by late 2030s | Breaking 2048-bit encryption |
| Cyber Insights 2025 Report | 2024 | 17%-34% chance by 2034, 79% by 2044 | Breaking RSA 2048 in 24 hours |
| Unisys | 2024 | 10 seconds | Perfect quantum computer |
| National Security Memorandum 10 (US) | 2022 | Full migration to PQC by 2035 | US Federal systems |
| Google Quantum AI | 2024 | Under 5 minutes for benchmark computation | Computation taking a supercomputer 10 septillion years |
| Craig Gidney and Martin Eker | 2021 | 8 hours | Using 20 million noisy qubits |
| Intro to Quantum.org | 2024 | ~ 20 million physical qubits, 8 hours | Factoring 2048-bit number, ~0.1% gate error |

Table 1 : Estimated Timeframe for Breaking RSA-2048

| Regulation/Treaty | Issuing Body/Region | Focus | Relevance to Financial Cybersecurity |
|---|---|---|---|
| Computer Fraud and Abuse Act (CFAA) | USA | Cybercrime | Prohibits unauthorized access to computer systems, relevant to AI and quantum-enabled attacks. |
| Gramm-Leach-Bliley Act (GLBA) | USA | Data Protection, Cybersecurity | Requires financial institutions to safeguard customer information, applicable to data security in the context of AI and quantum threats. |
| Sarbanes-Oxley Act (SOX) | USA | Financial Record Security, Cybersecurity | Includes cybersecurity components to ensure financial institutions address risks, relevant to AI and quantum impacts on financial data integrity. |
| Bank Secrecy Act (BSA) | USA | Anti-Money Laundering, Cybercrime | Requires controls to detect financial crime, which can be enhanced or threatened by AI and quantum technologies. |
| General Data Protection Regulation (GDPR) | European Union | Data Protection | Sets rules for processing personal data, relevant to data security and |

|  |  |  | privacy in the context of AI and potential quantum decryption. |
|---|---|---|---|
| Digital Operational Resilience Act (DORA) | European Union | Operational Resilience, Cryptography | Specifically addresses ICT risk management and cryptography in the financial sector, including mention of quantum risks. |
| Quantum Computing Cybersecurity Preparedness Act | USA | Quantum Cybersecurity | Mandates federal agencies to migrate to quantum-resistant cryptography, encouraging private sector adoption. |
| Budapest Convention on Cybercrime | Council of Europe | Cybercrime, International Cooperation | Aims to harmonize national laws and improve international cooperation in combating cybercrime, potentially applicable to AI-driven attacks. |
| UN Convention against Transnational Organized Crime | United Nations | Transnational Crime, Cybercrime | Addresses transnational organized crime, which can include cybercrime and may involve AI or quantum technologies in the future. |

**Table2 : Regulation/Treaty of Financial Cybersecurity**

| Feature | Quantum Key Distribution (QKD) | Post-Quantum Cryptography (PQC) |
|---|---|---|
| Security Principle | Based on the laws of quantum mechanics; any eavesdropping is detectable. | Based on mathematical problems believed to be hard for both classical and quantum computers. |
| Advantages | Unconditional security against eavesdropping, real-time detection of interception, potential for long-term security. | Implementable on existing classical infrastructure, standardization efforts underway, potentially more scalable for certain applications. |
| Disadvantages | Limited range due to photon loss, requires specialized hardware and infrastructure, higher implementation costs, authentication not inherent in standard protocols, potential physical layer vulnerabilities. | Security relies on the continued hardness of the underlying mathematical problems, may require more computational resources than current cryptography, ongoing research needed to ensure long-term security. |
| Maturity Level | More mature for key exchange, but challenges remain for long-distance and widespread deployment. | Standardization process recently completed for initial algorithms, implementation and testing are ongoing. |
| Cost | Generally higher due to specialized hardware and infrastructure requirements. | Potentially lower initial cost as it can leverage existing infrastructure, but performance optimization may require investment. |
| Complementary Role | Can be used for ultra-secure key exchange in critical communication | Can provide encryption and digital signatures for a wide range of |

| | links, complementing PQC for broader encryption needs. | applications, potentially securing data at rest and in transit across the financial ecosystem. |
|---|---|---|

**Table3 : Comparison Between Quantum Key Distribution (QKD) and Post-Quantum Cryptography (PQC)**

# References

1. The Role of AI and Cybersecurity in the Financial Sector - Software Mind, accessed March 15, 2025, https://softwaremind.com/blog/the-role-of-ai-and-cybersecurity-in-the-financial-sector/

2. Safeguarding central bank digital currency systems in the post-quantum computing age, accessed March 15, 2025, https://www.weforum.org/stories/2024/05/safeguarding-central-bank-digital-currency-systems-post-quantum-age/

3. Beyond Phishing: Exploring the Rise of AI-enabled Cybercrime - CLTC UC Berkeley Center for Long-Term Cybersecurity, accessed March 15, 2025, https://cltc.berkeley.edu/2025/01/16/beyond-phishing-exploring-the-rise-of-ai-enabled-cybercrime/

4. AI-powered cyberattacks are rising: 87% of businesses affected | bobsguide, accessed March 15, 2025, https://www.bobsguide.com/ai-powered-cyberattacks-are-rising-87-of-businesses-affected/

5. What Is Quantum Computing's Threat to Cybersecurity? - Palo Alto Networks, accessed March 15, 2025, https://www.paloaltonetworks.com/cyberpedia/what-is-quantum-computings-threat-to-cybersecurity

6. AI-Assisted Cyberattacks and Scams - NYU, accessed March 15, 2025, https://www.nyu.edu/life/information-technology/safe-computing/protect-against-cybercrime/ai-assisted-cyberattacks-and-scams.html

7. Most Common AI-Powered Cyberattacks | CrowdStrike, accessed March 15, 2025, https://www.crowdstrike.com/en-us/cybersecurity-101/cyberattacks/ai-powered-cyberattacks/

8. How Is Your Financial Institution Managing AI Cybersecurity Risks?, accessed March 15, 2025, https://www.ncontracts.com/nsight-blog/ai-cybersecurity-risks

9. AI Has Become an Integral Part of Fraud Prevention—and Fraud Attacks - PaymentsJournal, accessed March 15, 2025, https://www.paymentsjournal.com/ai-has-become-an-integral-part-of-fraud-prevention-and-fraud-attacks/

10. Banks must be wary of AI security risks, regulator says | Banking Dive, accessed March 15, 2025, https://www.bankingdive.com/news/banks-must-wary-ai-security-risks-says-regulator-nydfs/730186/

11. Some banks moving too slow to address AI-powered cyberthreats ..., accessed March 15, 2025, https://www.nextgov.com/cybersecurity/2024/03/some-banks-moving-too-slow-address-ai-powered-cyberthreats-treasury-says/395251/

12. Quantum is coming — and bringing new cybersecurity threats with it - KPMG International, accessed March 15, 2025, https://kpmg.com/xx/en/our-insights/ai-and-technology/quantum-and-cybersecurity.html

13. Quantum Computing - How it Changes Encryption as We Know It | Division of Information Technology - University of Maryland, accessed March 15, 2025, https://it.umd.edu/security-privacy-audit-risk-and-compliance-services-sparcs/topic-week/quantum-computing-how-it-changes-encryption-we-know-it

14. Criminals Use Generative Artificial Intelligence to Facilitate Financial Fraud, accessed March 15, 2025, https://www.ic3.gov/PSA/2024/PSA241203

15. OSFI-FCAC Risk Report - AI Uses and Risks at Federally Regulated Financial Institutions, accessed March 15, 2025, https://www.osfi-bsif.gc.ca/en/about-osfi/reports-publications/osfi-fcac-risk-report-ai-uses-risks-federally-regulated-financial-institutions

16. AI Fraud Detection: Preventing Scams with Generative AI - Veridas, accessed March 15, 2025, https://veridas.com/en/generative-ai-fraud/

17. AI-Powered Fraud Detection: All you need to know - Comidor, accessed March 15, 2025, https://www.comidor.com/blog/artificial-intelligence/ai-powered-fraud-detection/

18. The Role of AI in Modern Fraud Detection and Auditing - MindBridge, accessed March 15, 2025, https://www.mindbridge.ai/blog/the-role-of-ai-in-modern-fraud-detection-and-auditing/

19. Camouflage of AI in Cyber Crimes Vis-a Vis legal issues and Challenges,, accessed March 15, 2025, https://woxsen.edu.in/woxsen-law-review/wlr-papers/camouflage-of-AI-in-cyber-crimes-vis-a-vis-legal-issues-and-challenges/

20. Quantum computing and the financial system: opportunities and risks - Bank for International Settlements, accessed March 15, 2025, https://www.bis.org/publ/bppdf/bispap149.pdf

21. Impact of Quantum Computing on Finance Sector - CIO Influence, accessed March 15, 2025, https://cioinfluence.com/it-and-devops/impact-of-quantum-computing-on-finance-sector/

22. The Quantum Threat: How Financial Institutions Can Stay Ahead in an Evolving Cyber Landscape | WiCyS - Women in Cybersecurity, accessed March 15, 2025, https://www.wicys.org/the-quantum-threat-how-financial-institutions-can-stay-ahead-in-an-evolving-cyber-landscape/

23. Quantum computing in finance: A game-changer, eventually | ITWeb, accessed March 15, 2025, https://www.itweb.co.za/article/quantum-computing-in-finance-a-game-changer-eventually/xA9POvNEKe9qo4J8

24. Shor's Algorithm: Unlocking the Power of Quantum Computing | by Biraj karki | Medium, accessed March 15, 2025, https://medium.com/@birajkarki/shors-algorithm-unlocking-the-power-of-quantum-computing-13af3bdf9a7f

25. Quantum Cryptography - Shor's Algorithm Explained - Classiq, accessed March 15, 2025,

https://www.classiq.io/insights/shors-algorithm-explained

26. Preparing for Q-Day: Making payments quantum-safe, accessed March 15, 2025, https://thepaymentsassociation.org/article/preparing-for-q-day-making-payments-quantum-safe/

27. Researchers Say Here's How to Prepare Now For Post Quantum Cybersecurity, accessed March 15, 2025, https://thequantuminsider.com/2024/11/30/researchers-say-heres-how-to-prepare-now-for-post-quantum-cybersecurity/

28. Post-Quantum Computing Security Challenge - AIS Student Chapters - AIS Communities, accessed March 15, 2025, https://communities.aisnet.org/aisstudentchapters/eventshome/competitions/2025-competitions/postquantumcomputingchallenge

29. What is Quantum-Safe Cryptography? - IBM, accessed March 15, 2025, https://www.ibm.com/think/topics/quantum-safe-cryptography

30. Securing tomorrow: How NIST's post-quantum encryption standards will impact the financial sector - Moody's, accessed March 15, 2025, https://www.moodys.com/web/en/us/insights/quantum/how-nists-post-quantum-encryption-standards-will-impact-the-financial-sector.html

31. Cyber Insights 2025: Quantum and the Threat to Encryption - SecurityWeek, accessed March 15, 2025, https://www.securityweek.com/cyber-insights-2025-quantum-and-the-threat-to-encryption/

32. www.securityweek.com, accessed March 15, 2025, https://www.securityweek.com/cyber-insights-2025-quantum-and-the-threat-to-encryption/#:~:text=The%20timeline%20toward%20CRQC&text=In%202024%2C%20it%20estimated%20that,to%2079%25%20by%202044.%E2%80%9D

33. Toward a code-breaking quantum computer | MIT News, accessed March 15, 2025, https://news.mit.edu/2024/toward-code-breaking-quantum-computer-0823

34. The timelines: when can we expect useful quantum computers?, accessed March 15, 2025, https://introtoquantum.org/essentials/timelines/

35. China's Quantum Leap, and Why RSA Isn't at Risk (Yet) | Keysight Blogs, accessed March 15, 2025, https://www.keysight.com/blogs/en/tech/nwvs/2024/10/28/security-highlight-quantum-leap-in-china-and-why-rsa-isnt-at-risk-yet

36. How to enhance quantum security for the financial sector - The World Economic Forum, accessed March 15, 2025, https://www.weforum.org/stories/2024/11/pioneering-public-private-collaboration-financial-sector-secure-quantum-future/

37. Q-Day: Estimating and Preparing for Quantum Disruption in Cybersecurity | Secureworks, accessed March 15, 2025, https://www.secureworks.com/blog/predicting-q-day-and-impact-of-breaking-rsa2048

38. Quantum's Impact on Cybersecurity: The Hero and Villain - Viva Technology, accessed March 15, 2025, https://vivatechnology.com/news/quantum-s-impact-on-cybersecurity

39. Securing data in the post-quantum age - PwC, accessed March 15, 2025,

https://www.pwc.com/m1/en/publications/securing-data-in-the-post-quantum-age.html

40. What The Quantum Computing Cybersecurity Preparedness Act Means For National Security - Forbes, accessed March 15, 2025, https://www.forbes.com/councils/forbestechcouncil/2023/01/25/what-the-quantum-computing-cybersecurity-preparedness-act-means-for-national-security/

41. Singapore authority teams up with finance and tech firms on quantum security, accessed March 15, 2025, https://www.globalgovernmentfintech.com/singapore-monetary-authority-quantum-security-banks-tech-collaboration/

42. Securing the Future with Quantum-Safe Cryptography - Guidehouse, accessed March 15, 2025, https://guidehouse.com/insights/advanced-solutions/2025/quantum-safe-cryptography

43. Quantum Safe Financial Forum - Europol, accessed March 15, 2025, https://www.europol.europa.eu/cms/sites/default/files/documents/Quantum-safe-financial-forum-2025.pdf

44. Quantum Computing: The Urgent Need to Transition to Quantum-Resistant Cryptography - Bank Policy Institute, accessed March 15, 2025, https://bpi.com/quantum-computing-the-urgent-need-to-transition-to-quantum-resistant-cryptography/

45. Teetering on the brink of quantum utility – Not if, but when - Herbert Smith Freehills, accessed March 15, 2025, https://www.herbertsmithfreehills.com/insights/reports/2025/fsr-outlook-2025/teetering-on-the-brink-of-quantum-utility-not-if-but-when

46. Future outlook: The impact of quantum computing on financial services | London Daily News, accessed March 15, 2025, https://www.londondaily.news/future-outlook-the-impact-of-quantum-computing-on-financial-services/

47. How Post-Quantum Cryptography Impacts Financial Services' Cybersecurity - Redjack, accessed March 15, 2025, https://redjack.com/resources/quantum-computing-cybersecurity-financial-services

48. Examining Jurisdictional Issues in Cross-Border Financial Cybercrime in the US, accessed March 15, 2025, https://leppardlaw.com/federal/computer-crimes/examining-jurisdictional-issues-in-cross-border-financial-cybercrime-in-the-us/

49. Cyber Law: What You Need to Know - Axiom Law, accessed March 15, 2025, https://www.axiomlaw.com/guides/cyber-law

50. Cybercrime and legal frameworks | Criminal - Vocal Media, accessed March 15, 2025, https://vocal.media/criminal/cybercrime-and-legal-frameworks-sk91503yj

51. Worldwide Laws and Regulations Related to Cybercrime - zenarmor.com, accessed March 15, 2025, https://www.zenarmor.com/docs/network-security-tutorials/what-is-cybersecurity-laws-and-regulations

52. Top 9 Cybersecurity Regulations for Financial Services - UpGuard, accessed March 15, 2025, https://www.upguard.com/blog/cybersecurity-regulations-financial-industry

53. What global cyber and cybersecurity regulations are there? | CUBE, accessed March 15, 2025, https://cube.global/resources/compliance-corner/what-global-cyber-and-cybersecurity-regulations-are-there

54. How do U.S. cybersecurity and AI regulations influence international investigations of cybercrime in the financial sector? - PidginPLM, accessed March 15, 2025, https://www.pidginplm.com/2025/03/03/how-do-u-s-cybersecurity-and-ai-regulations-influence-international-investigations-of-cybercrime-in-the-financial-sector/

55. Financial Regulatory Agencies - Center for American Progress, accessed March 15, 2025, https://www.americanprogress.org/article/taking-further-agency-action-on-ai/financial-regulatory-agencies-chapter/

56. Cybersecurity Laws and Regulations Report 2025 USA - ICLG.com, accessed March 15, 2025, https://iclg.com/practice-areas/cybersecurity-laws-and-regulations/usa

57. CIOs must prepare their organizations today for quantum-safe cryptography - IBM, accessed March 15, 2025, https://www.ibm.com/think/insights/cios-must-prepare-their-organizations-today-for-quantum-safe-cryptography

58. What is the cyber security risk from quantum computing? - KPMG Australia, accessed March 15, 2025, https://kpmg.com/au/en/home/insights/2024/04/cyber-security-risk-from-quantum-computing.html

59. Quantum is coming — and bringing new cybersecurity threats with it - KPMG International, accessed March 15, 2025, https://kpmg.com/dp/en/home/insights/2024/04/quantum-and-cybersecurity.html

60. MAS Collaborates with Banks and Technology Partners on Quantum Security - Monetary Authority of Singapore, accessed March 15, 2025, https://www.mas.gov.sg/news/media-releases/2024/mas-collaborates-with-banks-and-technology-partners-on-quantum-security

61. www.centraleyes.com, accessed March 15, 2025, https://www.centraleyes.com/ai-regulation-in-finance/#:~:text=AI%20Governance%20in%20the%20United,growing%20focus%20on%20consumer%20protection.

62. Regulating AI in the financial sector: recent developments and main challenges, accessed March 15, 2025, https://www.bis.org/fsi/publ/insights63.htm

63. Managing Artificial Intelligence-Specific Cybersecurity Risks in the Financial Services Sector - Treasury Department, accessed March 15, 2025, https://home.treasury.gov/system/files/136/Managing-Artificial-Intelligence-Specific-Cybersecurity-Risks-In-The-Financial-Services-Sector.pdf

64. DOJ's Strategic Approach to Countering Cybercrime and AI Misuse - Lewis Brisbois, accessed March 15, 2025, https://lewisbrisbois.com/newsroom/legal-alerts/dojs-strategic-approach-to-countering-cybercrime-and-ai-misuse

65. State Laws Present Litigation Risks for Financial Industry's Artificial Intelligence Use | BCLP, accessed March 15, 2025, https://www.bclplaw.com/en-US/events-insights-news/state-laws-present-litigation-risks-for-financial-industrys-artificial-intelligence-use.html

66. Treasury's Post-2024 RFI Report on AI in Financial Services – Uses, Opportunities, and Risks | 01 | 2025 | Publications, accessed March 15, 2025, https://www.debevoise.com/insights/publications/2025/01/treasurys-post-2024-rfi-report-on-ai-in-financial

67. How to Manage AI-Specific Cybersecurity Risks in the Financial Services Sector, accessed March 15, 2025, https://www.cliffordchance.com/insights/resources/blogs/talking-tech/en/articles/2024/05/how-to-manage-ai-specific-cybersecurity-risks-in-the-financial-services-sector.html

68. AI Watch: Global regulatory tracker - Nigeria | White & Case LLP, accessed March 15, 2025, https://www.whitecase.com/insight-our-thinking/ai-watch-global-regulatory-tracker-nigeria

69. Overview of AI policy in 15 jurisdictions | Digital Watch Observatory, accessed March 15, 2025, https://dig.watch/updates/overview-of-ai-policy-in-15-jurisdictions

70. Cybercrime - Interpol, accessed March 15, 2025, https://www.interpol.int/Crimes/Cybercrime

71. S. Rept. 117-251 - QUANTUM COMPUTER CYBERSECURITY PREPAREDNESS ACT, accessed March 15, 2025, https://www.congress.gov/congressional-report/117th-congress/senate-report/251/1

72. H.R.7535 - 117th Congress (2021-2022): Quantum Computing Cybersecurity Preparedness Act, accessed March 15, 2025, https://www.congress.gov/bill/117th-congress/house-bill/7535

73. Quantum Cryptographic Algorithms for Securing Financial Transactions, accessed March 15, 2025, https://computerfraudsecurity.com/index.php/journal/article/view/32

74. Networking Feasibility of Quantum Key Distribution Constellation Networks - MDPI, accessed March 15, 2025, https://www.mdpi.com/1099-4300/24/2/298

75. How Quantum Computing Benefits Financial Services [2025] - SpinQ, accessed March 15, 2025, https://www.spinquanta.com/news-detail/how-quantum-computing-benefits-financial-services20250219023634

76. The Positive and the Negative Impacts of Quantum Computers on the Finance Sector, accessed March 15, 2025, https://cloudsecurityalliance.org/articles/the-positive-and-the-negative-impacts-of-quantum-computers-on-the-finance-sector

77. Classical vs. Quantum Cryptography: Strengths and Challenges - Juniper Research, accessed March 15, 2025, https://www.juniperresearch.com/resources/infographics/classical-vs-quantum-cryptography-strengths-and-challenges/

78. Quantum Key Distribution: A Viable Solution for Businesses? - BizTech Magazine, accessed March 15, 2025, https://biztechmagazine.com/article/2025/03/quantum-key-distribution-qkd-perfcon

79. Challenges and Opportunities in Quantum Cryptography - EMB Global, accessed March 15, 2025, https://blog.emb.global/challenges-and-opportunities-in-quantum-cryptography/

80. Quantum Key Distribution (QKD) Market Size, Trends & Forecast - Verified Market Research, accessed March 15, 2025, https://www.verifiedmarketresearch.com/product/quantum-key-distribution-qkd-market/

81. Challenges in Implementing Quantum Cryptography and How to Overcome Them, accessed March 15, 2025, https://www.coherentmarketinsights.com/blog/challenges-in-implementing-quantum-cryptography-and-how-to-overcome-them-1226

82. Quantum Key Distribution Market Challenges and Opportunities, accessed March 15, 2025, https://www.coherentmarketinsights.com/market-insight/quantum-key-distribution-market-5971/market-challenges-and-opportunities

83. Challenges of implementing quantum key distribution - HLK - IP, accessed March 15, 2025, https://www.hlk-ip.com/news-and-insights/challenges-of-implementing-quantum-key-distribution/

84. The Invisible Threat: How Quantum Computing Could Break Today's Encryption? | by Akitra, accessed March 15, 2025, https://medium.com/@akitrablog/the-invisible-threat-how-quantum-computing-could-break-todays-encryption-888e3ea99cf3

85. New quantum safe cryptographic standards future proofing financial security in the quantum age - Capgemini, accessed March 15, 2025, https://www.capgemini.com/insights/expert-perspectives/new-quantum-safe-cryptographic-standards-future-proofing-financial-security-in-the-quantum-age/

86. Why the financial sector is embracing a dual quantum-safe strategy - ID Quantique, accessed March 15, 2025, https://www.idquantique.com/why-the-financial-sector-is-embracing-a-dual-quantum-safe-strategy/

87. Post-Quantum Cryptography: The Future of Secure Communications and the Role of Standards - AppViewX, accessed March 15, 2025, https://www.appviewx.com/blogs/post-quantum-cryptography-the-future-of-secure-communications-and-the-role-of-standards/

88. G7 Cyber Expert Group Statement On Planning For The Opportunities And Risks Of Quantum Computing, accessed March 15, 2025, https://home.treasury.gov/system/files/136/G7-CYBER-EXPERT-GROUP-STATEMENT-PLANNING-OPPORTUNITIES-RISKS-QUANTUM-COMPUTING.pdf

89. G7 Cyber Expert Group Recommends Action to Combat Financial Sector Risks from Quantum Computing | U.S. Department of the Treasury, accessed March 15, 2025, https://home.treasury.gov/news/press-releases/jy2609

90. G7 Cyber Expert Group recommends action to combat financial sector risks from quantum computing | Global Regulation Tomorrow, accessed March 15, 2025, https://www.regulationtomorrow.com/eu/g7-cyber-expert-group-recommends-action-to-combat-financial-sector-risks-from-quantum-computing/

91. Call for action: urgent plan needed to transition to post-quantum cryptography together, accessed March 15, 2025, https://www.europol.europa.eu/media-press/newsroom/news/call-for-action-urgent-plan-needed-to-transition-to-post-quantum-cryptography-together

92. 10 seconds to break: Preparing for quantum security threats - Unisys, accessed March 15, 2025, https://www.unisys.com/blog-post/cis/10-seconds-to-break-preparing-for-quantum-security-threats/

93. Quantum Computing and the Financial Sector | Publications | Cleary Gottlieb, accessed March 15, 2025, https://www.clearygottlieb.com/news-and-insights/publication-listing/quantum-computing-and-the-financial-sector-world-economic-forum-lays-out-roadmap-towards-quantum-security

94. Quantum Technology for Securing Financial Messaging | QED-C, accessed March 15, 2025,

https://quantumconsortium.org/financial24/

95. Artificial intelligence and the challenge for global governance, accessed March 15, 2025, https://agatadata.com/wp-content/uploads/2024/11/AI-the-Challenge-for-Global-Governance-06_24-Digital-Society-Initiative.pdf

96. NIST recommends timelines for transitioning cryptographic algorithms | PQShield, accessed March 15, 2025, https://pqshield.com/nist-recommends-timelines-for-transitioning-cryptographic-algorithms/